\documentclass[10pt, conference]{IEEEtran}
\IEEEoverridecommandlockouts

\usepackage{cite}
\usepackage{multicol}

\usepackage{amssymb}

\usepackage{amsmath}
\usepackage{amsfonts}
\usepackage[T1]{fontenc}
\usepackage{algpseudocode}
\usepackage[utf8]{inputenc}
\usepackage[english]{babel}
\usepackage[usenames, dvipsnames]{color}
\usepackage{pifont}
\usepackage{booktabs}
\usepackage{mathtools}
\usepackage{textcomp}
 \usepackage{graphicx}
\usepackage{subfigure}
\usepackage[ruled,vlined,commentsnumbered]{algorithm2e}
\usepackage{amsthm}
\usepackage{setspace}

\SetKwInput{KwIn}{Inputs}

\theoremstyle{definition}

\theoremstyle{remark}

\theoremstyle{plain}

\newcommand{\RNum}[1]{\uppercase\expandafter{\romannumeral #1\relax}}
\newcommand{\forallK}{\forall\,k \in \mathcal{K}}

\def\BibTeX{{\rm B\kern-.05em{\sc i\kern-.025em b}\kern-.08em
    T\kern-.1667em\lower.7ex\hbox{E}\kern-.125em}}

\begin{document}

\title{Optimal Weight Scheme for Fusion-Assisted Cooperative Multi-Monostatic Object Localization in 6G Networks}
\author{\IEEEauthorblockN{ Maximiliano Rivera Figueroa, Pradyumna Kumar Bishoyi, and Marina Petrova}
\IEEEauthorblockA{Mobile Communications and Computing, RWTH Aachen University, Aachen, Germany \\
Email: \{maximiliano.rivera, pradyumna.bishoyi, petrova\}@mcc.rwth-aachen.de}
}

\maketitle

\begin{abstract}
Cooperative multi-monostatic sensing enables accurate positioning of passive targets by combining the sensed environment of multiple base stations (BS).
In this work, we propose a novel fusion algorithm that optimally finds the weight to combine the time-of-arrival (ToA) and angle-of-arrival (AoA) likelihood probability density function (PDF) of multiple BSs. In particular, we employ a log-linear pooling function that fuses all BSs' PDFs using a weighted geometric average. We formulated an optimization problem that minimizes the Reverse Kullback–Leibler Divergence (RKLD) and proposed an iterative algorithm based on the Monte Carlo importance sampling (MCIS) approach to obtain the optimal fusion weights. Numerical results verify that our proposed fusion scheme with optimal weights outperforms the existing benchmark in terms of positioning accuracy in both unbiased (line-of-sight only) and biased (multipath-rich environment) scenarios.

\end{abstract}

\section{Introduction} \label{sec:Introduction}
The forthcoming sixth-generation (6G) wireless standard is expected to facilitate new emerging services such as extended reality (XR), autonomous vehicle tracking, digital twins, and smart cities, which require a tight integration of communication and positioning. To cater to these services' demands, integrated sensing and communication (ISAC) is emerging as a key technology enabler to empower the existing cellular base stations (BSs) with sensing capabilities to gather additional information about the surrounding active and passive objects.  

A recent report by the ITU-R \cite{ITU2022} examines the potential of cooperative ISAC networks to support ultra-precise positioning services by coordinating the sensing operation of multiple BSs and processing the information. 
In this context, a few recent works \cite{figueroa2023cooperative, Device-Free2022,figueroa2023jcs}, present the idea of \textit{cooperative multi-monostatic sensing}, where first each BS performs monostatic sensing and thereafter the BSs' estimates  are fused for passive target localization. The key benefit of this cooperative sensing scheme is to improve target localization accuracy by exploiting the diversity of each BS in terms of its geometry or location, field-of-view (FoV), and the physical channel between the BS and the target. Nevertheless, the localization performance depends not solely on the sensing accuracy of each BS but also on the fusion mechanism that combines the sensing measurements from different BSs. Clearly, a BS with the worst channel condition will have erroneous sensing measurement, contributing negatively to the localization performance. Therefore, our main focus in this work is designing an effective fusion scheme for the cooperative multi-monostatic sensing system that addresses the challenges described above, which remains an interesting open problem.  

Localization schemes can use different types of measurements, such as time-of-arrival (ToA), time-difference-of-arrival (TDoA), received signal strength (RSS), angle-of-arrival (AoA), and their hybrid combinations. For example, authors in \cite{adi_survey, Guvenc2009, figueroa2023cooperative} use the ToA-based range measurement technique for target localization. The ToA-based localization scheme is more effective when the target is in line-of-sight (LOS) with respect to the BS. However, its accuracy and precision decrease where the multipath component arises due to the non-line-of-sight (NLOS) conditions. In our previous work \cite{figueroa2023cooperative}, we illustrate the effect of NLOS paths on the target localization and also discuss three different fusion-based estimation methods to improve localization performance in a multipath-rich environment. To tackle the effect of NLOS condition, authors in \cite{Naseri2019,Henninger2022} employ a hybrid ToA-AoA approach for target localization. Authors in \cite{Naseri2019} formulate a joint ToA-AoA maximum likelihood estimation (MLE) problem to minimize the localization error. The joint MLE function is a weighted sum of ToA and AoA error distributions, where the weights are the respective variances. Further, authors in \cite{Henninger2022} formulate an optimization problem aimed at maximizing the joint log-likelihood functions of ToA and AoA. They consider the error due to multipath as outliers and propose an iterative weight assignment algorithm to remove it. Nevertheless, the proposed algorithm is only suitable for active uplink localization. 
All these above works suffer from the limitation of utilizing i) either a constant value for the weights assigned to combine ToA and AoA information that is determined by the variance (both ToA and AoA) of the observations, which consequently is specific to the scenario and can not be generalized or ii) a sub-optimal value that results in increased localization error after fusing.

  

In this paper, we present a cooperative multi-monostatic sensing-based passive target localization scheme that optimally calculates the fusion weights and thereafter fuses the hybrid ToA-AoA information of each participating BS. In the first step, we model ToA and AoA error likelihood functions of each BS as Gaussian distribution and Von Mises-Fisher (VMF) distribution, respectively. Each BS's ToA and AoA error probability density functions (PDFs) are parameterized by their relative location to the target and propagation environment. In the second step, all the PDFs from BSs are fused at the central processing unit (CPU). Specifically, we envision this as a \textit{probabilistic opinion pooling problem} \cite{Koliander2022} where the CPU employs a pooling function, i.e. a geometric average function, in order to fuse the PDFs from all BSs, generating an \textit{aggregated} PDF. 
The CPU aims to fuse the PDFs so that the resultant PDF is as close as possible to the "ground truth" PDF, which is the optimal manner of fusing the ToA and AoA information.
Developing an appropriate weight assignment algorithm that minimizes the difference between aggregated PDF and ground truth PDF presents a significant challenge.
For that, we consider the Reverse Kullback–Leibler Divergence (RKLD) as a metric measuring the discrepancy between two PDFs \cite{Wymeersch2009}. 
Further, we formulate an optimization problem that minimized the RKLD by finding the optimal weights in an iterative manner. Subsequently, we utilize the obtained weights to compute the optimal location of the target.
The main contributions of our work are
\begin{itemize}
    \item \textit{Fusion mechanism}: We introduce a novel fusion scheme based on a pooling function that incorporates each BS's ToA and AoA measurements, which are key factors in deciding the localization accuracy. We formulate an optimization problem that minimizes the RKLD to obtain the optimal weights. Due to the difficulty in obtaining a closed-form solution for the optimal solution, we employ Monte Carlo Importance Sampling (MCIS) method to transform the objective function. Thereafter, we apply the Newton-Raphson algorithm to obtain the optimal weights in an iterative manner.
    \item \textit{Performance evaluation}: To show the efficacy of the proposed fusion scheme, we evaluate the positioning error in both \textbf{unbiased }(LOS path only) and \textbf{biased} (multipath-rich environment) scenarios and compare our proposed weight assignment with equal weight scenario. We show that our fusion scheme outperforms the equal weight scheme in terms of positioning accuracy in both scenarios. We also give insights about the importance of using an \textit{importance sampling} approach and the number of samples needed to reach a desired level of accuracy.
\end{itemize}

The rest of the paper is organized as follows. We describe the system model in Section \ref{sec:SystemModel}. In Section \ref{sec:PowerAllocation}, we describe the formulated optimization problem and explain the proposed algorithm in \ref{sec:Proposedsoln}. Further, we show the performance evaluation in Section \ref{sec:PerformanceEvaluation} and conclude the paper in Section \ref{sec:Conclusions}.
\section{System Model}\label{sec:SystemModel}
We consider a network consisting of $K$ BSs in an urban area where each BS has full duplex capabilities and operates in Frequency Range 1 (FR1) band. Each BS is a monostatic radar, and the network aims to localize passive targets that lie in the sensing coverage given by the FoV of the BSs. We assume that all the BSs are synchronized and connected to a CPU via a high-capacity backhaul link. The position of each BS is assumed to be known and is denoted as $\boldsymbol{x}_k, \,k \in \mathcal{K}$, with $\mathcal{K} = \{1, \dots, K\}$. The position of each target to be estimated is unknown and is denoted as $\boldsymbol{x}$.

Each BS is able to estimate the distance and its relative angle to the target independently by sending an orthogonal frequency division multiplexing (OFDM) signal and collecting the echoes. Similar to \cite{figueroa2023cooperative}, the distance is estimated using the periodogram algorithm. The BS estimates the AoA using the multiple signal classification (MUSIC) algorithm \cite{Pucci2022}. 
Then, the estimates from all BSs are shared with the CPU, whose aim is to fuse the estimates optimally. Henceforth, we consider an observation model to represent the distance and AoA estimates.
\subsection{Observation model}\label{sec:obsSyst}
In the observation model, the distance and angle information are assumed to be noisy.
For the distance observations, we follow the model proposed in \cite{Huang2010}, in which the estimated distance ($\hat{d}_k$) between the target and the $k$-th BS is given by
\begin{equation}
    \hat{d}_k = d_k +b_k + n_k, \label{eq:distObservations}
\end{equation}
where $d_k$ is the true distance, $b_k$ is the bias due to multipath, which depends on the propagation environment, and $n_k$ is the measured noise, with respect to the $k$-th BS. The estimated distance is assumed to be obtained through ToA measurements. Therefore, distance observations and ToA observations are used interchangeably. The bias term $b_k$ can be modelled as an exponential decay random variable, uniform, or delta distribution \cite{Huang2010}. The bias in the estimation is mainly contributed by the ground reflection \cite{Ge2023}, and it can be as strong as the LOS signal. The measured noise is modelled as a zero-mean normal random variable where $n_k \sim \mathcal{N}(0, \sigma_k^2)$, with $\sigma_k$ being the standard deviation of the $k$-th BS's distance observations.

For the AoA observations, we follow the model proposed in \cite{Naseri2019}, where the azimuth $\hat{\varphi}_k$ and elevation $\hat{\theta}_k$ AoA, at the $k$-th BS, are given by
\begin{align}
\hat{\varphi}_k &= \varphi_k + w_{k,\,\varphi}\\
\hat{\theta}_k &= \theta_k + w_{k,\,\theta}
\end{align}
where $\varphi_k$ and $\theta_k$ are the true azimuth and true elevation of the target, respectively, and $w_{k,\,\varphi}$ and $w_{k,\,\theta}$ are the measured noise for the azimuth and elevation AoA, respectively. Note that $w_{k,\,\varphi}$ and $w_{k,\,\theta}$ can be modelled as a zero-mean normal distribution, where $w_{k,\,\varphi} \sim \mathcal{N}(0, \sigma_{k,\,\varphi}^2)$ and $w_{k,\,\theta} \sim \mathcal{N}(0, \sigma_{k,\,\theta}^2)$, with $\sigma_{k,\,\varphi}$ and $\sigma_{k,\,\theta}$ the standard deviation of the azimuth and elevation AoA observations of the $k$-th BS, respectively.  Furthermore, it is assumed that the observations of distance and AoA are independent, and its variances can be accurately estimated either by computing them from the observations or by approximating them through performance bounds \cite{Naseri2019} \footnote{It is worth noting that the presented system model can be expanded to bistatic scenarios in which the receiver is not collocated with the transmitter. Applying the proposed solution to these networks is straightforward.}.

\subsection{Likelihood Functions}
In order to model the observation errors, we use different likelihood functions to model the ToA and AoA errors. The distance errors for the BS $k \in \mathcal{K}$ are modelled as Gaussian distribution as in \cite{figueroa2023cooperative}:
\begin{align}
    q_{k,\text{ToA}}(\boldsymbol{x}|\hat{d}_k) = \frac{1}{\sqrt{2\pi \sigma_k^2 }}  \exp \Big[\frac{-(\hat d_k - ||\boldsymbol{x}_k - \boldsymbol{x}||)^2}{2\,\sigma_k^2} \Big].
    \label{eq:qToA}
\end{align}
On the other hand, the AoA errors are modelled by the Von Mises-Fisher (VMF) distribution, which is the equivalent of the Gaussian distribution but in a two-dimension sphere space \cite{Mardia1975}. The likelihood function of the VMF distribution for the BS $k \in \mathcal{K}$ is given by 
\begin{align}
    q_{k,\,\text{AoA}}(\boldsymbol{u}_k| \hat{\boldsymbol{u}}_k, \kappa_k) = \frac{1}{2\pi I_0( \kappa_k)} \exp\Big(\kappa_k
\, \hat{\boldsymbol{u}}_k^T \boldsymbol{u}_k \Big),
\label{eq:qVMF}
\end{align}
where $\boldsymbol{u}_k=(\boldsymbol{x} - \boldsymbol{x}_k) / (||\boldsymbol{x} - \boldsymbol{x}_k||)$ is a unitary vector pointing to the true position of the target and $\hat{\boldsymbol{u}}_k$ is a unitary vector pointing to the observed direction, which is a function of $\hat{\varphi}_k$ and $\hat{\theta}_k$. $I_0(\cdot)$ is the modified Bessel function of the first kind of order zero, and $\kappa_k$ is the concentration parameter of the $k$-th BS, which is similar to the variance of a Gaussian distribution. 

It is important to note that our analysis is applicable to both single-target and multiple-target scenarios, as the target's position is estimated based on the observation model of each BS. In a multi-target scenario, the number of targets is initially determined by employing an additional association stage, which relies on a conventional algorithm such as density-based spatial clustering of applications with noise (DBSCAN) \cite{DBSCAN1996} or data association algorithm described in \cite{Device-Free2022}.

Once the observations are obtained, each BS shares its tuple of estimate, i.e. $(\hat{d}_k,\,\hat{\varphi}_k, \hat{\theta}_k)$, with the CPU. The CPU fuses the tuples from all the $K$ BSs to enhance the estimation accuracy using error distributions, namely $q_{k, \text{ToA}}$ and $q_{k, \text{AoA}}$. 
The fusion mechanism plays a crucial role in situations when the BSs' observations are subject to bias and noise. Each BS's observation varies based on factors such as its FoV and its location relative to the target. Therefore, ensuring the optimal fusion of different observations is of the utmost importance. Failing to do so leads to an increase in the positioning error. Consequently, the following section presents an optimal way to fuse different observations.

\section{Problem Formulation}\label{sec:PowerAllocation}
Finding the optimal way to fuse the observations of different BSs given a set of conditions can be referred to as an opinion pool problem \cite{Koliander2022}. This problem assumes that the observation of each BS is based on a probability density function (PDF), which is shared with the CPU. The CPU uses a pooling function to combine all the PDFs optimally and build the called \textit{aggregate} PDF. Let $q_k(\mathbf{\upsilon})$ be the PDF used at the $k$-th BS, where $\boldsymbol{\upsilon}$ is a multivariate random variable, and $g(\cdot)$ the pooling function that combines the PDFs of the opinion pool $\boldsymbol{q} = [q_1,\dots,q_K]^T$, then the aggregate PDF is given by
\begin{equation}
    q(\boldsymbol{\upsilon}) = g(q_1,\dots,q_K)(\boldsymbol{\upsilon}).
\end{equation}
The optimal pooling function when the observed data is independent is given by the log-linear pooling function \cite{Koliander2022}:
\begin{equation}
    g(q_1,\dots,q_K)(\boldsymbol{\upsilon}) = \frac{\prod_{k=1}^K q_k^{w_k}(\boldsymbol{\upsilon}) }{\int \prod_{k=1}^K q_k^{w_k}(\boldsymbol{\zeta})d\boldsymbol{\zeta}}, \label{eq:log-linearPool}
\end{equation}
where $w_k$ is the weight given to the $k$-th BSs, such that $\sum_{k=1}^K w_k = 1$. In order to avoid Eq.~\eqref{eq:log-linearPool} being undefined, the opinion pool $\boldsymbol{q}$ has to have positive PDFs on their domains. 
The CPU aims to choose a set of $\{w_k\}_{k=1}^K$ such that the aggregate PDF is equal to the \textit{ground truth} PDF $p(\boldsymbol{\upsilon})$, which fuse the observations optimally.
The PDF $p(\boldsymbol{\upsilon})$ is usually unknown and challenging to obtain. Therefore, in order to evaluate how close the aggregate PDF $q(\boldsymbol{\upsilon})$ is to the optimal PDF $p(\boldsymbol{\upsilon})$, we use the Reverse Kullback–Leibler Divergence (RKLD) \cite{Koliander2022, Wymeersch2009}. The RKLD is a discrepancy measure that assesses the similarity of two PDFs, and it is given by
\begin{equation}
    \mathcal{D}_{KL} (q||p) = \int_{\Upsilon} p(\boldsymbol{\upsilon}) \log\Bigg(\frac{p(\boldsymbol{\upsilon})}{q(\boldsymbol{\upsilon})}\Bigg) d\boldsymbol{\upsilon}, \label{eq:KLD}
\end{equation}
where $\Upsilon$ is the domain of the aggregate PDF. It can be noticed that $\mathcal{D}_{KL} (q||p)$ is minimized when $q(\boldsymbol{\upsilon})\to p(\boldsymbol{\upsilon})$.
Further, substituting Eq.~\eqref{eq:log-linearPool} into Eq.~\eqref{eq:KLD}, and doing some algebraic manipulation, Eq.~\eqref{eq:KLD} is then written as
\begin{align}\label{dkl_mid}
    \mathcal{D}_{KL} (q||p) = \sum_{k=1}^K\,w_k\,\mathcal{D}_{KL} (q_k||p) + C - I_{\boldsymbol{w}}(\boldsymbol{q}),
\end{align}
where $C$ is a constant value independent of $q(\boldsymbol{\upsilon})$ given by
\begin{align}
    C = (K-1) \mathcal{D}_{KL}(1||p)
\end{align}
and 
\begin{equation}
    I_{\boldsymbol{w}}(\boldsymbol{q}) = -\log \int_{\Upsilon} \prod_{k=1}^K q_k^{w_k}(\boldsymbol{\upsilon}) \,d\boldsymbol{\upsilon},
\end{equation}
with $\boldsymbol{w} = [w_1,\dots,w_K]^T$ the weight's vector. $I_{\boldsymbol{w}}(\boldsymbol{q})$ is referred as the generalized Chernoff information (GCI) of $\boldsymbol{q}$ with parameter $\boldsymbol{w}$ \cite{Lehrer2019}.

From Eq.~\eqref{dkl_mid}, it can be observed that as the GCI increases, the value of RKLD decreases, and the aggregate PDF is closer to the ground truth PDF based on the measure used. Hence, maximizing the GCI leads to the minimum RKLD value possible when the ground truth PDF is unknown. Therefore, an optimization problem is formulated as follows

\begin{align}
    \max_{\boldsymbol{w}} \quad & I_{\boldsymbol{w}}(\boldsymbol q) & \tag{P1} \label{eq:P1}\\
    \textrm{s.t.} \quad & \sum_{k=1}^K w_k = 1\tag{P1.a}\\
	\quad & w_k\geq 0, \quad \forall\, k \in \mathcal{K} \tag{P1.b}
\end{align}
Notice that the optimal weights $\boldsymbol{w}^*$ are given by the argument that maximizes \ref{eq:P1}. Furthermore, by considering that $\boldsymbol{q} = [q_{1,\,\text{ToA}}, \,q_{1,\,\text{AoA}}, \dots, q_{K,\,\text{ToA}}, \,q_{K,\,\text{AoA}}]^T$, \ref{eq:P1} can be transformed into the following optimization problem

\begin{align}
    \min_{\boldsymbol{w}} \quad & \int_{\Upsilon} \prod_{k=1}^K q_{k,\,\text{ToA}}^{w_{k,\,\text{ToA}}}(\boldsymbol{\upsilon}) \cdot q_{k,\,\text{AoA}}^{w_{k,\,\text{AoA}}}(\boldsymbol{\upsilon}) \,d\boldsymbol{\upsilon} & \tag{P2} \label{eq:P2} \\
    \textrm{s.t.} \quad & \sum_{k=1}^K (w_{k,\,\text{ToA}}+ w_{k,\,\text{AoA}}) = 1\tag{P2.a}\label{eq:P2a}\\
	\quad & w_{k,\,\text{ToA}}\geq 0, \;\;w_{k,\,\text{AoA}}\geq 0, \quad \forall\, k \in \mathcal{K} \tag{P2.b}
\end{align}

\noindent in which the logarithmic function can be removed as it is a monotone function. Note that $\boldsymbol{w} = [w_{1,\,\text{ToA}}, \,w_{1,\,\text{AoA}}, \dots, w_{K,\,\text{ToA}}, \,w_{K,\,\text{AoA}}]^T$. 

Once the optimal weights $\boldsymbol{w}^*$ are obtained and used in Eq.~\eqref{eq:log-linearPool} the optimal position of the target can be computed, and it can be solved using the maximum likelihood algorithm \cite{figueroa2023cooperative}. The optimization problem can be written as
\begin{equation}
    \hat{\boldsymbol{x}} = \arg \max_{\boldsymbol{x}} \prod_{k=1}^K q_{k,\,\text{ToA}}^{w_{k,\,\text{ToA}}^*}(\boldsymbol{x}|\hat{d}_k) \cdot q_{k,\,\text{AoA}}^{w_{k,\,\text{AoA}}^*}(\boldsymbol{u}_k| \hat{\boldsymbol{u}}_k, \kappa_k) \tag{P3} \label{eq:P3}
\end{equation}
where $\boldsymbol{u}_k$ depends on the azimuth and elevation of $\boldsymbol{x}$ referred to the $k$-th BS. Hence, solving \ref{eq:P1} and using optimal weights to solve \ref{eq:P3} leads to the optimal target's position. 

\section{Proposed Solution}\label{sec:Proposedsoln}
In this section, we present our proposed fusion scheme for determining the optimal target position. First, we propose an iterative-based method to determine the optimal weights and thereafter solve the Problem \ref{eq:P3} using the optimal weights to find the optimal position. 
\subsection{Optimal weights estimation}
It can be shown that the objective function of \ref{eq:P2} is convex by employing Hölder's inequality of order one \cite{boydconvex}, and the constraint set of Problem \ref{eq:P2} is compact and convex. Thus, the Problem \ref{eq:P2} is convex in nature. However, solving it is non-trivial and does not have a closed-form solution \cite{Lehrer2019}. One way to solve it is by employing grid-based searching methods, which are computationally costly. Especially for a dimension higher than two, it is almost intractable \cite{Ahmed2012}. 
Another way of solving \ref{eq:P2} is by particle approximation methods, where a set of particles approximates the PDFs. We follow a similar approach as the method proposed in \cite{Ahmed2012}, where a Monte Carlo Importance Sampling (MCIS) is used. This method relies on finding an importance sampling (IS) PDF $\psi(\boldsymbol{\upsilon})$ that samples the PDFs of the opinion pool $\boldsymbol{q}$. In order to do that, $N_{s}$ samples are drawn from $\psi(\boldsymbol{\upsilon})$, namely $\{\boldsymbol{\upsilon}_{\iota}\}_{\iota=1}^{N_s}\sim \psi(\boldsymbol{\upsilon})$, an used to sample the opinion pool $\boldsymbol{q}$.

In order to reach the optimal point when using the MCIS method, the IS PDF has to fulfil three conditions \cite{Lehrer2019}. First, $\psi(\boldsymbol{\upsilon})$ is often chosen to minimizes the variance of the sampled log-linear pooling function. Second, it has to have a similar peak structure to the sampled function and a similar support domain. Third, the number of samples drawn has to be sufficiently large to be close to an optimal solution.

Thereupon, in order to build the MCIS method under the conditions stated above, the objective function of \ref{eq:P2} is rewritten as

\begin{align}
&\int_{\Upsilon} \prod_{k=1}^K q_{k,\,\text{ToA}}^{w_{k,\,\text{ToA}}}(\boldsymbol{\upsilon}) \cdot q_{k,\,\text{AoA}}^{w_{k,\,\text{AoA}}}(\boldsymbol{\upsilon}) \,d\boldsymbol{\upsilon} \nonumber \\
&= \int_{\Upsilon} \psi(\boldsymbol{\upsilon})\frac{\prod_{k=1}^{K} q_{k,\,\text{ToA}}^{w_{k,\,\text{ToA}}}(\boldsymbol{\upsilon}) \cdot q_{k,\,\text{AoA}}^{w_{k,\,\text{AoA}}}(\boldsymbol{\upsilon}) }{\psi(\boldsymbol{\upsilon})} \,d\boldsymbol{\upsilon} \nonumber \\
&= \sum_{\iota=1}^{N_s} \frac{\prod_{k=1}^{K} q_{k,\,\text{ToA}}^{w_{k,\,\text{ToA}}}[\boldsymbol{\upsilon_{\iota}}] \cdot q_{k,\,\text{AoA}}^{w_{k,\,\text{AoA}}}[\boldsymbol{\upsilon_{\iota}}]}{\psi[\boldsymbol{\upsilon_{\iota}}]},\label{eq:MCIS_opt}
\end{align} 

where $\psi(\boldsymbol{\upsilon})$ sample the division of the log-linear pooling function over the IS PDF. As stated in \cite{Ahmed2012}, a good choice of $\psi(\boldsymbol{\upsilon})$ is given by the same sampled function, that is
\begin{equation}
\psi(\boldsymbol{\upsilon}) = \prod_{k=1}^{K} q_{k,\,\text{ToA}}^{w_{k,\,\text{ToA}}^{IS}}(\boldsymbol{\upsilon}) \cdot q_{k,\,\text{AoA}}^{w_{k,\,\text{AoA}}^{IS}}(\boldsymbol{\upsilon}) \label{eq:ISpdf}
\end{equation}
where $w_{k,\,\text{ToA}}^{IS}$ and $w_{k,\,\text{AoA}}^{IS}$ are the weights for the IS PDF and are defined as equal weights, such that $w_{k,\,\text{ToA}}^{IS} = w_{k,\,\text{AoA}}^{IS} = 1/(2K),\forallK$. 
Notice that, in this manner, the sampling function possesses a similar peak structure as the sampled function.
Then, once the samples of $\psi(\boldsymbol{\upsilon})$ are drawn, and as the objective function of \ref{eq:P2} is convex, we utilize the damped Newton-Raphson algorithm to solve the optimization problem iteratively. 
Newton-Raphson algorithm is used as its convergence is rapid in general, and its behaviour is well-defined when the problem scales to multiple dimensions \cite{boydconvex}.

\begin{algorithm}
	\small
	\SetKwInOut{Input}{Inputs}
	\SetKwInOut{Output}{Outputs}
	\Input{$\hat{\boldsymbol{d}}$, $\hat{\boldsymbol{\varphi}}$, $\hat{\boldsymbol{\theta}}$, $\boldsymbol{w}^{IS}$, $N_s$, $\epsilon$}
	\Output{ $ \hat{\boldsymbol{x}}$}
    Build vector $\boldsymbol{q}$ using $\hat{\boldsymbol{d}}$, $\hat{\boldsymbol{\varphi}}$, $\hat{\boldsymbol{\theta}}$\\ 
    Compute IS PDF $\psi(\boldsymbol{\upsilon})$ using \eqref{eq:ISpdf}\\
    Draw $N_s$ samples $\{\boldsymbol{\upsilon}_{\iota}\}_{\iota=1}^{N_s}\sim \psi(\boldsymbol{\upsilon})$\\
    Compute IS PDF $\psi[\boldsymbol{\upsilon}_{\iota}],\iota \in [1,\dots, N_s]$\\
    Sample $\boldsymbol{q}$ with $\{\boldsymbol{\upsilon}_{\iota}\}_{\iota=1}^{N_s}$\\
	Set $ converge=0, i=0$, and $\boldsymbol{w}^i \leftarrow \boldsymbol{w}^{IS}$\\
	\While{converge = 0}{
        Find $\boldsymbol{w}^{i+1}$ by using Newton-Raphson algo. in Eq.~\eqref{eq:MCIS_opt}\\
        Normalize the weights $\boldsymbol{w}^{i+1}$\\
        $i \leftarrow i+1$\\
		\uIf{$||\boldsymbol{w}^{i} - \boldsymbol{w}^{i-1}||<\epsilon$}
        { $\boldsymbol{w}^{*} \leftarrow \boldsymbol{w}^{i}$ \\
        $converge \leftarrow 1$}
	}
    Get $\hat{\boldsymbol{x}}$ by solving \ref{eq:P3} using $\boldsymbol{w}^{*}$ and sampled $\boldsymbol{q}$
	\caption{Position estimation using MCIS}
	\label{algo:MCIS}
\end{algorithm}
\vspace{-0.25cm}
\subsection{Finding the Optimal Position}
After obtaining the optimal weights, the next step is to solve the Problem \ref{eq:P3} to obtain the optimal position. Given the MCIS method already yields the aggregate PDF and solving Problem \ref{eq:P2} provides the optimal weight ($\boldsymbol{w}^{*}$), we can determine the optimal position of the target ($\hat{\boldsymbol{x}}$) by searching over all the possible values of the sampled aggregate PDF. Note that the aggregate PDF has a bounded support, which is a combination of the support of each PDF. This results in a reduced search space.

Our complete solution is presented in Algorithm~\ref{algo:MCIS}, where $\boldsymbol{\hat{d}} = [\hat{d}_1,\dots,\hat{d}_K]$, $\boldsymbol{\hat{\varphi}} = [\hat{\varphi}_1,\dots,\hat{\varphi}_K]$, $\boldsymbol{\hat{\theta}} = [\hat{\theta}_1,\dots,\hat{\theta}_K]$,  $\boldsymbol{w}^{IS} = [w_{1,\,\text{ToA}}^{IS}, w_{1,\,\text{AoA}}^{IS},\dots,w_{K,\,\text{ToA}}^{IS},w_{K,\,\text{AoA}}^{IS}]$, and $\epsilon$ is a pre-defined threshold.
In the algorithm, as the constraint \eqref{eq:P2a} is relaxed, we add an extra step of normalization.
\subsection{Complexity Analysis}

Let $Y_1$ be the number of iterations needed for the Newton-Raphson algorithm to converge. Then, the overall complexity of Algorithm~\ref{algo:MCIS}  is given by $\mathcal{O}(Y_1 N_s^3 (2K)^4 )$, where $2K$ is the number of PDFs used in the opinion pool ($\boldsymbol{q}$). The higher complexity of the algorithm arises due to the computation of the Hessian matrix in the Newton-Raphson algorithm. Furthermore, as the number of BSs ($K$) able to be fused is limited in outdoor scenarios, the sampling number ($N_s$) dominates the algorithm's complexity. Therefore, restricting the value of the sampling number is essential to contain the complexity of the proposed algorithm. 

\section{Performance Evaluation}\label{sec:PerformanceEvaluation}
\vspace{-0.05cm}
\subsection{Setup and Scenarios}\label{sec:SetupScenario}

\begin{figure}[t]

    \centering
    \includegraphics[width=0.8\linewidth]{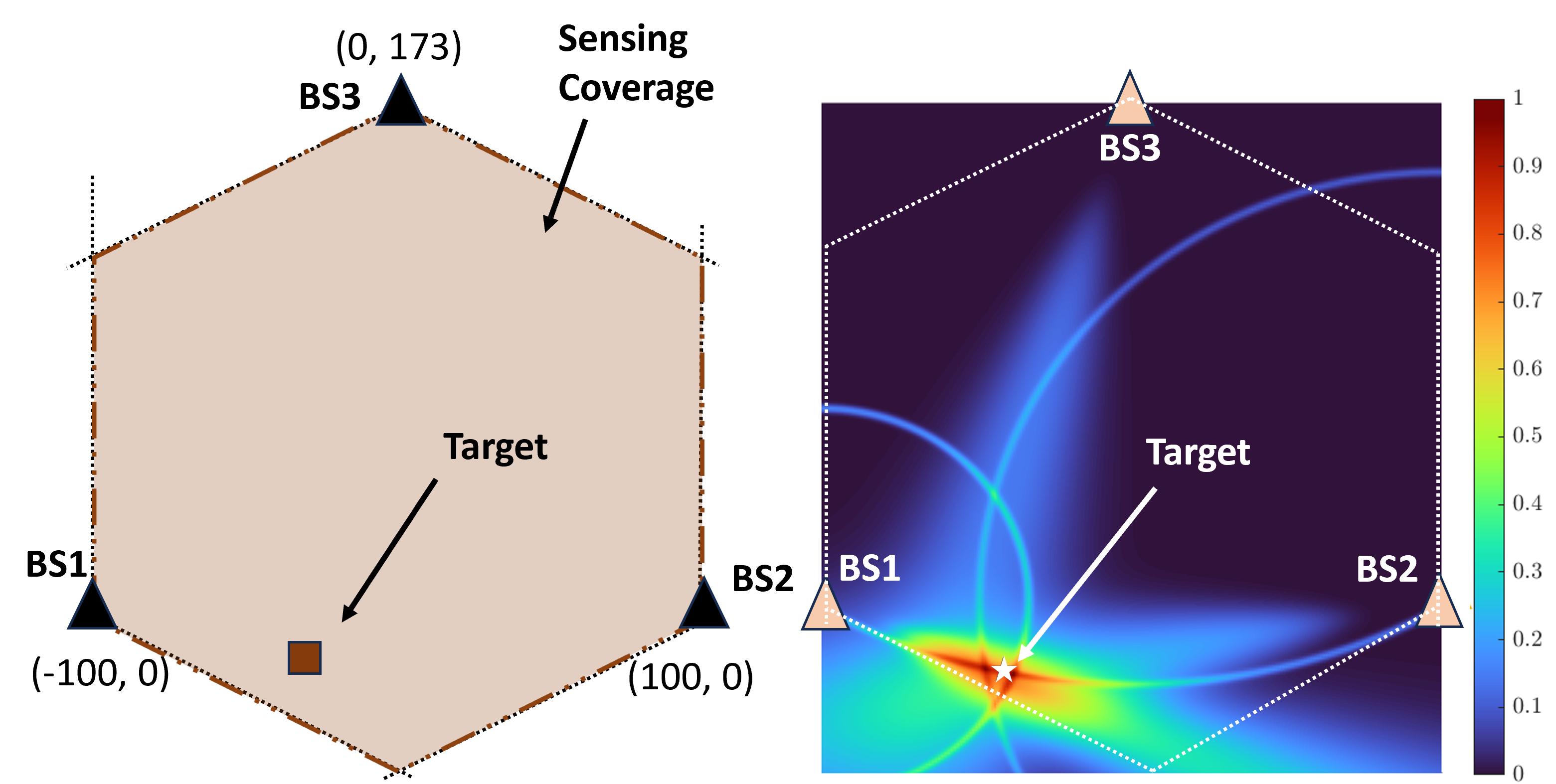}
    \caption{Illustration of the scenario for an arbitrary target's position. The figure on the left depicts the scenario and the sensing coverage, whereas, the figure on the right depicts the normalized error PDFs for ToA and AoA observations for the given target's position.}
    \label{fig:scenario}
    \vspace{-0.35cm}
\end{figure}

To evaluate the performance of the proposed algorithm, we use the setup illustrated in Fig.~\ref{fig:scenario}. The figure on the left illustrates our evaluation scenario with three BSs, a target, and the sensing coverage inside the hexagon. The figure on the right shows the likelihood function of ToA and AoA PDFs. Values close to 1 indicate a higher probability of the target's presence. We perform Monte Carlo simulations to assess the performance of the proposed algorithm. In each iteration, the target is placed randomly within the sensing coverage. 
The BSs are placed at a height of $10$ m, whereas the target has a height of $1$ m. 

In order to demonstrate the efficacy of the proposed fusion scheme, we evaluate the positioning error in two different scenarios. In the first scenario, the distance observations are unbiased, that is, $b_k = 0,\,\forallK$. Thus, we assume LOS channel from the BSs to the target. In the second scenario, the distance observations are biased, with $b_k$ following a uniform distribution. 

In the ToA observation model, the noise variance of the distance observations is set to $\sigma_k = 1\,\text{m},\,\forallK$, a typical value when the bandwidth used is $100$ MHz \cite{figueroa2023jcs}. The bias term is modelled as a uniform distribution with $b_k \sim \mathcal{U}(0, 5]\,\text{m}$. The azimuth and elevation AoA noise variances are set to $\sigma_{k,\,\varphi} = \sigma_{k,\,\theta} = 3.2^{\circ},\,\forallK$. Further, in the VMF distribution, we use a concentration parameter of $\kappa = 10$ \cite{Henninger2022}. In order to get accurate results, we use a sampling number of $N_x = N_y = 1000$.
Finally, the convergence threshold is set to $\epsilon = 10^{-4}$. 
To assess the performance of the proposed algorithm, we compare the positioning error using optimal weights (OW) given by Algorithm~\ref{algo:MCIS} versus equal weights (EW), which is a widely used benchmark \cite{Naseri2019}.

\subsection{Simulation Results and Discussion}\label{sec:SimResults}

\begin{figure*}[t]
\centering
	\subfigure[]{
		\includegraphics[width=156pt, trim={2.5cm 6cm 3.6cm 7cm},clip]{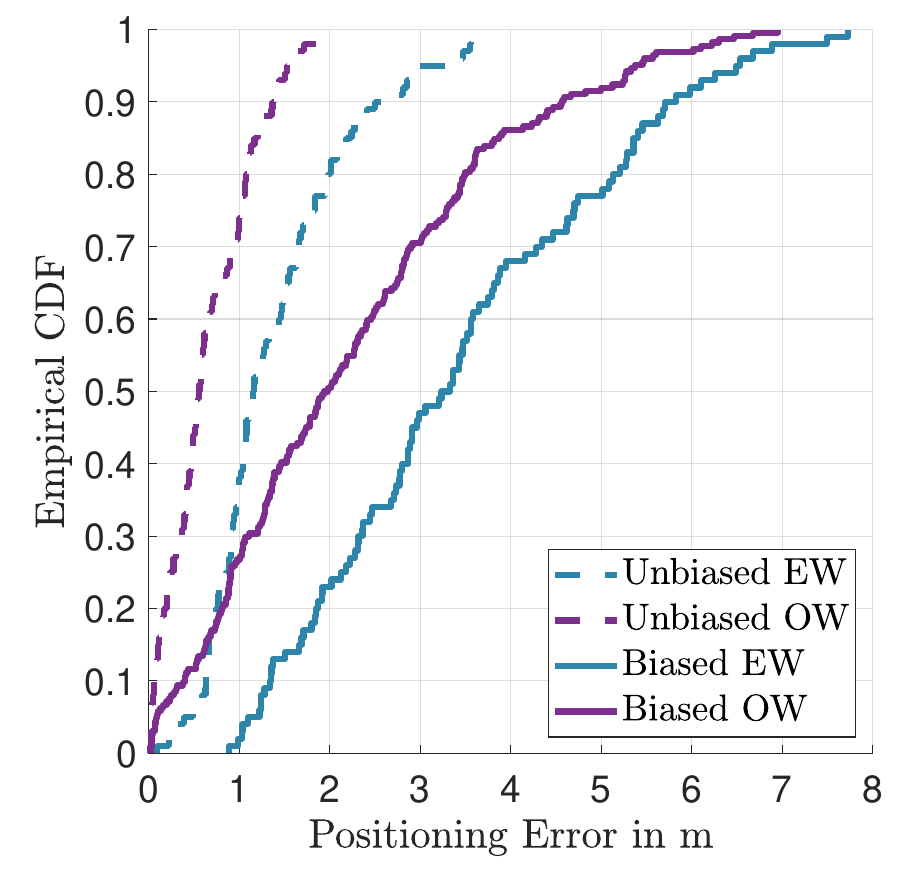}
    \label{fig:ToA_biased_v_unbiased}}
	\hfill
	\subfigure[]{ 
\includegraphics[width=156pt, trim={2.5cm 6cm 3.6cm 7cm},clip]{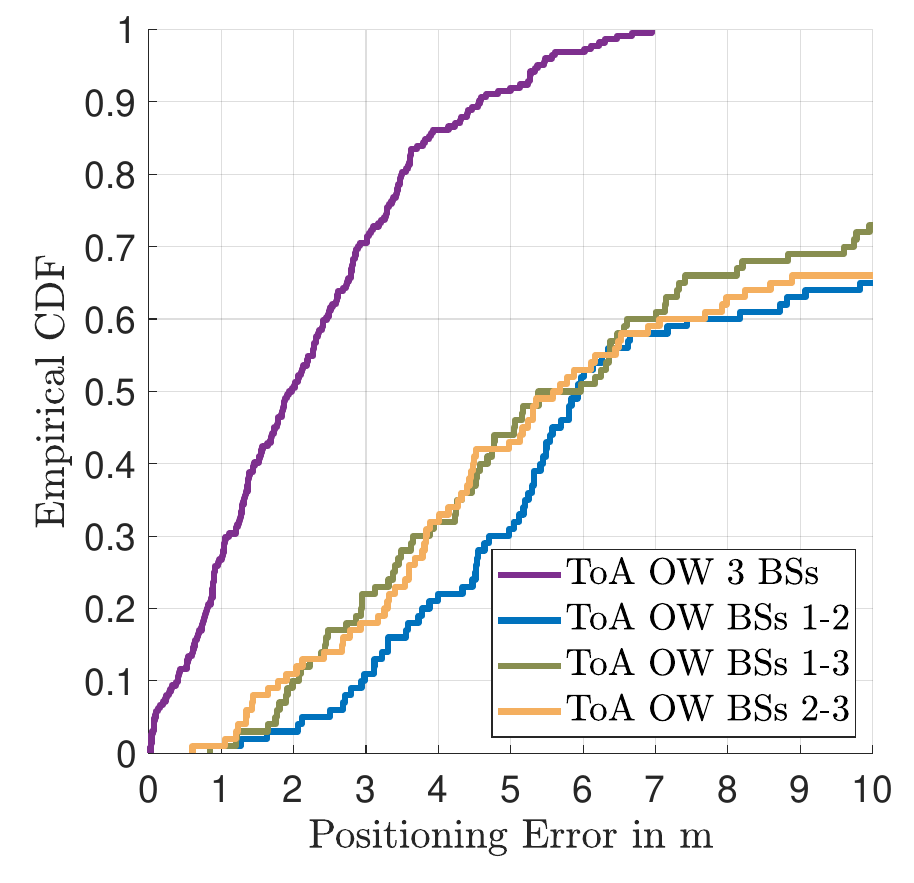}
    \label{fig:ToA_numberBS_comparison}
	}
	\hfill
	\subfigure[]{ 
    \includegraphics[width=156pt, trim={2.5cm 6cm 3.6cm 7cm},clip]{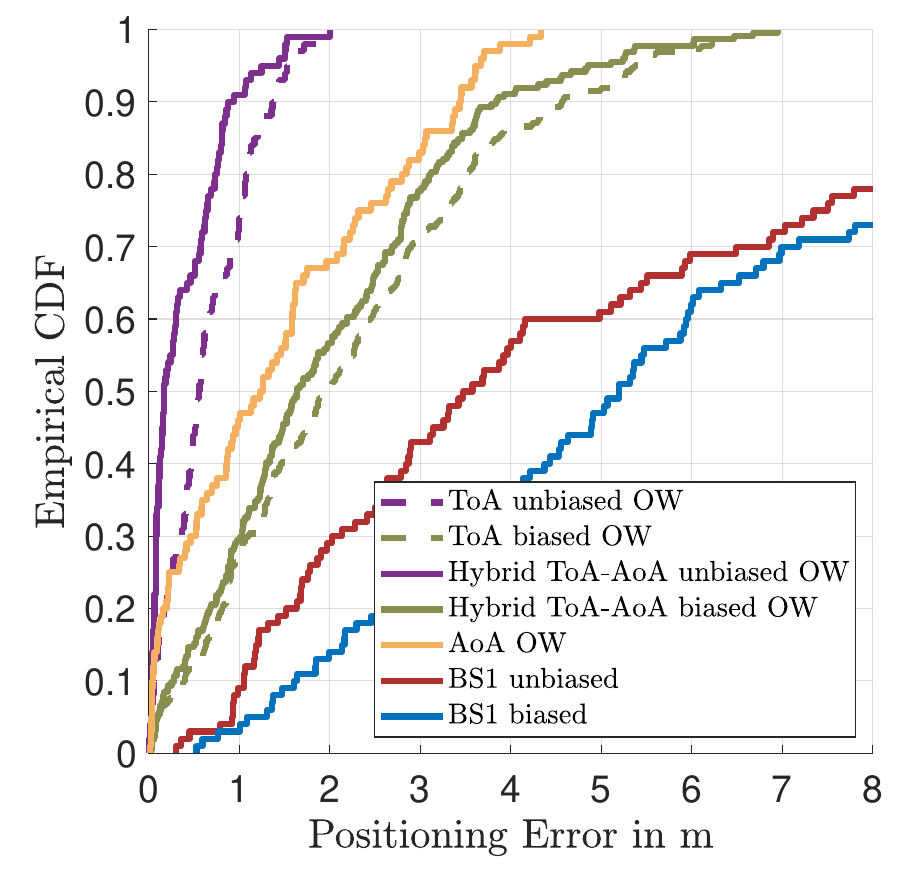}
    \label{fig:pE_ToA_AoA}
	}
	\caption{From left to right: (a) Positioning error comparison between equal weights (EW) and optimal weights (OW) for unbiased and biased distance observations, (b) Comparison of positioning error for three BSs' distance fusion versus two BSs' distance fusion. The fusion of two BSs' observations is presented over all possible combinations, (c) Positioning error comparison for ToA, AoA, and hybrid ToA-AoA fusion, and single BS estimation.}
	\label{fig:Distance_K2}
    \vspace{-0.4cm}
\end{figure*}



\begin{figure}[t]
    \centering
    \includegraphics[width=0.65\linewidth, trim={1.5cm 6.5cm 1.5cm 7cm},clip]{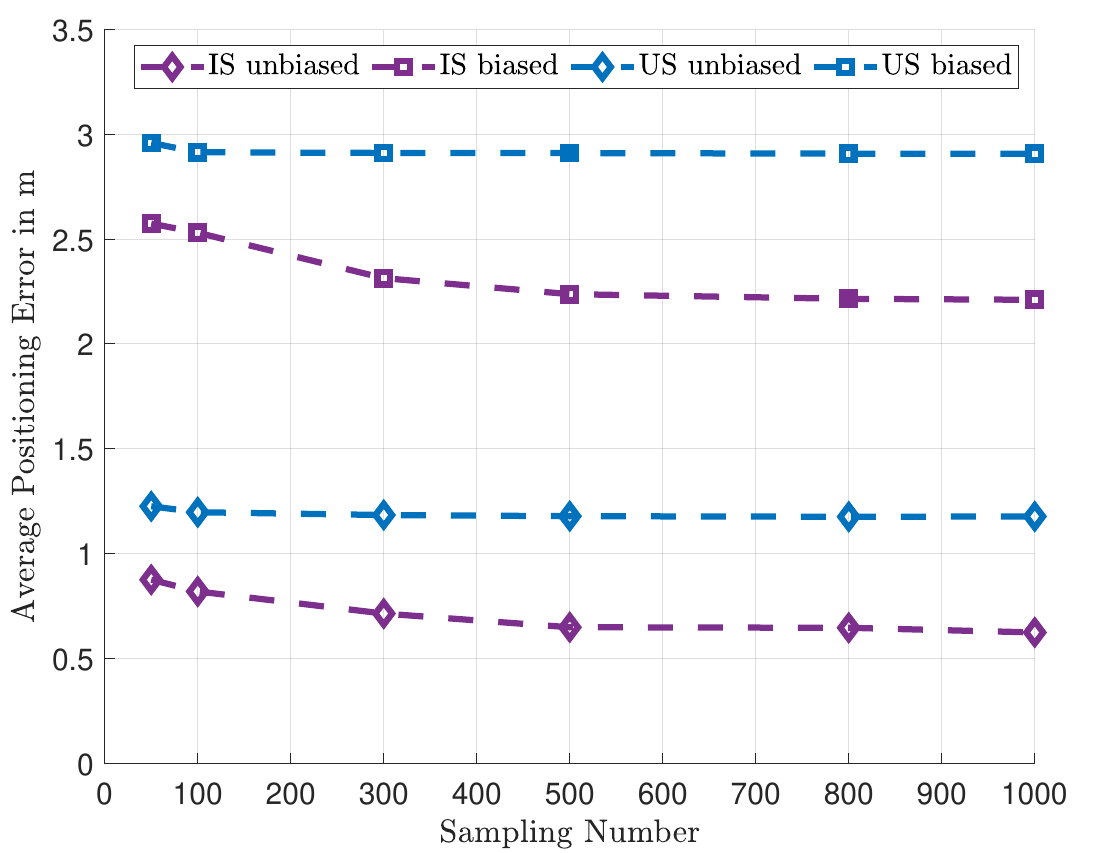}
    \caption{Comparison between Importance Sampling (IS) and Uniform Sampling (US).}
    \label{fig:IS_v_US}
    \vspace{-0.5cm}
\end{figure}

Fig.~\ref{fig:ToA_biased_v_unbiased} depicts the positioning error's cumulative density function (CDF) for both biased and unbiased scenarios when using equal and optimal weights for distance observations. Fusing unbiased distances leads to higher accuracy than biased observations. 
Additionally, as the proposed algorithm minimizes the RKLD, bringing the aggregate PDF closer to its ground truth, a significant improvement in accuracy can be achieved in both scenarios. Using OW in the unbiased scenario achieves a position error of $1$ m or less for $80\%$ of the cases. In the biased scenario, a position error of only $3.5$ m is achieved at the same confidence level.

Fig.~\ref{fig:ToA_numberBS_comparison} illustrates the effect of the number of participating BSs on the overall positioning error. We compare the performance of fusing ToA observation of two BSs and three BSs in the biased scenario, i.e. in the presence of multipath. Note that the fusion is done based on the proposed OW-based algorithm. 
It can be clearly observed that adding one BS more (from two to three) brings a high gain in accuracy. 

Next, we analyze the benefit of fusing both ToA and AoA information, as opposed to fusing only ToA or AoA information from all three BSs. Similar to other figures, we compare the performance of the proposed algorithm for both biased and unbiased scenarios. In Fig.~\ref{fig:pE_ToA_AoA}, it can be seen that 
using both ToA-AoA observations (Hybrid ToA-AoA) and fusing with optimal weights we achieve slightly better positioning performance than only using ToA in both scenarios. For example, in the unbiased scenario, our scheme exhibits the position error of $1$ m $90\%$ of the cases. Furthermore, we observe that while the fusion of only AoA information exhibits lower accuracy compared to the unbiased ToA case, it outperforms the Hybrid ToA-AoA fusion in the biased scenario. This highlights the impact of biased observations, which displace the true distance and, therefore, alter the optimal point in the Problem \ref{eq:P3}. Finally, for the hybrid ToA-AoA-based positioning, we assess the error when there is only one BS available. Without loss of generality, we include BS1's position estimation for our analysis. The results clearly show that any kind of fusion (either ToA or AoA only or hybrid ToA-AoA) with three BSs overpasses the position estimation of individual BSs. This clearly emphasizes that fusion with optimal weight is a crucial factor for enhancing positioning estimation performance.

In Fig.~\ref{fig:IS_v_US}, we assess the MCIS algorithm's performance by comparing it against a uniform sampling PDF for different sizes of drawn samples $N_x$, with $N_y = N_x$. In both scenarios, the uniform sampling PDF fails to improve its performance when increasing the number of samples, giving a constant average positioning error regardless of the number of samples drawn. We also observe that the MCIS improves its performance with the increase of the sampling number, reaching the optimal point for a sampling number of $500$ in both scenarios. 
\section{Conclusions}\label{sec:Conclusions}
This paper proposes an algorithm that optimally fuses the ToA and AoA observations when the ground truth PDF is unknown. We determine the optimal weights for a geometric weighted average in the fusion stage by minimizing the RKLD using a Monte Carlo Importance Sampling method. Our results exhibit the effectiveness of the proposed algorithm where the optimal weights outperform the benchmark scheme even in a multipath-rich environment. Furthermore, we demonstrate that the positioning accuracy improves when fusing hybrid ToA-AoA information of multiple BSs, as opposed to the individual estimation of a single BS. 
Finally, we give insights about the importance of using an importance sampling approach and the number of samples needed to reach a desired level of accuracy.

\section*{Acknowledgment}
This work was partially funded by the Federal Ministry of Education and Research Germany within the project ”KOMSENS-6G” under grant 16KISK128 and partially funded under the Excellence Strategy of the Federal Government and the Länder under grant "RWTH-startupPD 441-23".
\bibliographystyle{IEEEtran}
\bibliography{Fusion_based.bib}

\begin{thebibliography}{10}
\providecommand{\url}[1]{#1}
\csname url@samestyle\endcsname
\providecommand{\newblock}{\relax}
\providecommand{\bibinfo}[2]{#2}
\providecommand{\BIBentrySTDinterwordspacing}{\spaceskip=0pt\relax}
\providecommand{\BIBentryALTinterwordstretchfactor}{4}
\providecommand{\BIBentryALTinterwordspacing}{\spaceskip=\fontdimen2\font plus
\BIBentryALTinterwordstretchfactor\fontdimen3\font minus \fontdimen4\font\relax}
\providecommand{\BIBforeignlanguage}[2]{{%
\expandafter\ifx\csname l@#1\endcsname\relax
\typeout{** WARNING: IEEEtran.bst: No hyphenation pattern has been}%
\typeout{** loaded for the language `#1'. Using the pattern for}%
\typeout{** the default language instead.}%
\else
\language=\csname l@#1\endcsname
\fi
#2}}
\providecommand{\BIBdecl}{\relax}
\BIBdecl

\bibitem{ITU2022}
``{REPORT ITU-R M.2516-0 - Future technology trends of terrestrial International Mobile Telecommunications systems towards 2030 and beyond},'' \url{https://www.itu.int/dms\_pub/itu-r/opb/rep/R-REP-M.2516-2022-PDF-E.pdf}, 11 2022, (Accessed on 11/18/2023).

\bibitem{figueroa2023cooperative}
M.~R. Figueroa, P.~K. Bishoyi, and M.~Petrova, ``{Cooperative Multi-Monostatic Sensing for Object Localization in 6G Networks},'' \emph{arXiv preprint arXiv:2311.14591}, 2023.

\bibitem{Device-Free2022}
{Q. Shi\textit{et al.}}, ``{Device-Free Sensing in OFDM Cellular Network},'' \emph{IEEE J. Sel. Areas Commun.}, vol.~40, no.~6, pp. 1838--1853, 2022.

\bibitem{figueroa2023jcs}
M.~R. Figueroa, P.~K. Bishoyi, and M.~Petrova, ``{Power Allocation Scheme for Device-Free Localization in 6G ISAC Networks},'' \emph{arXiv preprint arXiv:2402.10660}, 2024.

\bibitem{adi_survey}
S.~Aditya, A.~F. Molisch, and H.~M. Behairy, ``{A Survey on the Impact of Multipath on Wideband Time-of-Arrival Based Localization},'' \emph{Proc. IEEE}, vol. 106, no.~7, pp. 1183--1203, 2018.

\bibitem{Guvenc2009}
I.~Guvenc and C.-C. Chong, ``{A Survey on TOA Based Wireless Localization and NLOS Mitigation Techniques},'' \emph{IEEE Commun. Surveys Tuts.}, vol.~11, no.~3, pp. 107--124, 2009.

\bibitem{Naseri2019}
H.~Naseri and V.~Koivunen, ``{A Bayesian Algorithm for Distributed Network Localization Using Distance and Direction Data},'' \emph{IEEE Trans. Signal Inf. Process. Netw.}, vol.~5, no.~2, pp. 290--304, 2019.

\bibitem{Henninger2022}
{M. Henninger \textit{et al.}}, ``{Probabilistic 5G Indoor Positioning Proof of Concept with Outlier Rejection},'' in \emph{Proc. IEEE EuCNC/6G Summit}, 2022, pp. 249--254.

\bibitem{Koliander2022}
{G. Koliander \textit{et al.}}, ``{Fusion of Probability Density Functions},'' \emph{Proc. IEEE}, vol. 110, no.~4, pp. 404--453, 2022.

\bibitem{Wymeersch2009}
H.~Wymeersch, J.~Lien, and M.~Z. Win, ``{Cooperative Localization in Wireless Networks},'' \emph{Proc. IEEE}, vol.~97, no.~2, pp. 427--450, 2009.

\bibitem{Pucci2022}
L.~Pucci, E.~Paolini, and A.~Giorgetti, ``{System-Level Analysis of Joint Sensing and Communication Based on 5G New Radio},'' \emph{{IEEE J. Sel. Areas Commun.}}, vol.~40, no.~7, pp. 2043--2055, 2022.

\bibitem{Huang2010}
{J. Y. Huang \textit{et al.}}, ``{Dilution of precision for mobile location in Non-Line-of-Sight environments},'' in \emph{Proc. IEEE WICON}, 2010, pp. 1--8.

\bibitem{Ge2023}
{Y. Ge \textit{et al.}}, ``{Analysis of V2X Sidelink Positioning in sub-6 GHz},'' in \emph{Proc. IEEE JC\&S}, 2023, pp. 1--6.

\bibitem{Mardia1975}
K.~V. Mardia, ``{Statistics of Directional Data},'' \emph{J. R. Stat. Soc.: Series B (Methodological)}, vol.~37, no.~3, pp. 349--371, 1975.

\bibitem{DBSCAN1996}
{M.Ester \textit{et al.}}, ``{A density-based algorithm for discovering clusters in large spatial databases with noise},'' ser. KDD'96.\hskip 1em plus 0.5em minus 0.4em\relax AAAI Press, 1996, p. 226–231.

\bibitem{Lehrer2019}
N.~Lehrer, O.~Tslil, and A.~Carmi, ``{Log-linear Chernoff Fusion for Distributed Particle Filtering},'' in \emph{Proc. IEEE FUSION}, 2019, pp. 1--9.

\bibitem{boydconvex}
S.~Boyd and L.~Vandenberghe, \emph{{Convex optimization}}.\hskip 1em plus 0.5em minus 0.4em\relax Cambridge university press, 2004.

\bibitem{Ahmed2012}
N.~R. Ahmed and M.~Campbell, ``{Fast Consistent Chernoff Fusion of Gaussian Mixtures for Ad Hoc Sensor Networks},'' \emph{IEEE Trans. Signal Process.}, vol.~60, no.~12, pp. 6739--6745, 2012.

\end{thebibliography}

\end{document}